\documentclass[aip,pof,amsmath,amssymb]{revtex4-1}
\usepackage{graphicx, epstopdf}
\usepackage{dcolumn}
\usepackage{color}
\usepackage{bm}
\usepackage{hyperref}
\usepackage{multirow}
\begin{document}
\title{Precession-driven dynamos in a full sphere and the role of large scale cyclonic vortices}
\author{Yufeng Lin}
\email{yl552@cam.ac.uk}
\affiliation{Institute of Geophysics, ETH Zurich, Sonneggstrasse 5, 8092 Zurich, Switzerland }
\affiliation{Department of Applied Mathematics and Theoretical Physics, University of Cambridge, Wilberforce Road, 
CB3 0WA, Cambridge, UK}
\author{Philippe Marti}
 \affiliation{Department of Applied Mathematics, University of Colorado Boulder, Boulder,
Colorado 80309, USA}
\author{Jerome Noir}
\author{Andrew Jackson}
\affiliation{Institute of Geophysics, ETH Zurich, Sonneggstrasse 5, 8092 Zurich, Switzerland }


\begin{abstract}
Precession has been proposed as an alternative power source for planetary dynamos. Previous hydrodynamic simulations suggested that precession can generate very complex flows in planetary liquid cores [Y. Lin, P. Marti, and J. Noir, ``Shear-driven parametric instability in a precessing sphere," Physics of Fluids \textbf{27}, 046601 (2015)]. In the present study, we numerically investigate the magnetohydrodynamics of a precessing sphere. We demonstrate precession driven dynamos in different flow regimes, from laminar to turbulent flows. In particular, we highlight the magnetic field generation by large scale cyclonic vortices, which has not been explored previously. In this regime, dynamos can be sustained at relatively low Ekman numbers and magnetic Prandtl numbers, which paves the way for planetary applications.
\end{abstract}
\maketitle
\section{Introduction} \label{sec:Intro}
The Earth's magnetic field, which is believed to be generated by fluid motions in the outer core through the so-called geodynamo mechanism, has been in existence for at least 3 billion years according to paleomagnetic records. \cite{Jones2007Treatise, Tarduno2007} It is thought that the geodynamo is powered by compositional and thermal convection in the outer core.\cite{Jones2007Treatise} However, this conventional view of the geodynamo is called into question because of a tight energy budget. \cite{Nimmo2007} In particular, recently revised estimates of thermal conductivity that are higher than previously thought have placed the convection geodynamo in a more restricted position.\citep{Pozzo2012} Alternatively, Bullard \cite{Bullard1949} first proposed that precession, a change of the orientation of the rotation axis, is a potential power source to generate the Earth's magnetic field. From an energetic point of view, precession driven \textit{laminar} flow cannot extract sufficient energy to maintain the Earth's magnetic field. \cite{Loper1975,Rochester1975} In contrast, turbulent flows driven by precession can dissipate much more energy and thus are possible to sustain the geomagnetic field.\cite{Malkus1968,Kerswell1996}  In addition to the geodynamo, it has been proposed that the ancient lunar dynamo may be sustained by precession.  \cite{Dwyer2011,LeBars2011Nature} However, these studies did not take into account the constraints of realistic magnetohydrodynamics (MHD) driven by the precessional forcing. 

Gans \cite{Gans1971} first experimentally  studied MHD in a precessing cylinder filled with liquid sodium, showing the signature of amplified magnetic 
fields but ultimately no self-sustained dynamo action. In the last decade, several numerical simulations have demonstrated that precession-driven flows can sustain magnetic fields through dynamo action in spherical, \cite{Tilgner2005,Tilgner2007a} spheroidal \cite{Wu2009,Hullermann2013} and cylindrical \cite{Nore2011,Cappanera2016} geometries.  In contrast with laboratory experiments and planetary cores, numerical simulations usually  adopted a much higher ($P_m \geqslant 1$) magnetic Prandtl number $P_m$ (the ratio of the kinematic viscosity to the magnetic diffusivity) than that is appropriate for liquid metals. Therefore, the simulations are generally dominated by viscous dissipation rather than the Ohmic dissipation. In addition, due to limited computational resources, numerical models use a relatively large Ekman number ($E\geqslant 10^{-4}$) which measures the typical ratio between the viscous force and the Coriolis force. The present numerical study aims at shedding light on precession driven dynamos at relatively low Ekman numbers and magnetic Prandtl numbers. 

We work in a full sphere geometry. In this geometry only viscous coupling to the boundary is possible, and topographic couples that are effective in cylindrical and spheroidal geometries are entirely absent. Based on our hydrodynamic simulations, \cite{Lin2015} we investigate precession driven dynamos in different flow regimes. It is of particular interest that the nonlinear evolution of precessional instabilities can lead to a few dominant cyclonic vortices (i.e. rotating in the same direction as the background rotation), which are elongated along the rotation axis of the fluid. \cite{Mouhali2012,Lin2015} Hereafter we refer to these vortices as large scale cyclonic vortices (LSCV). These large scale vortices are thought to be a  favourable flow structure for magnetic field generation.  \cite{Guervilly2015} Indeed, our numerical simulations suggest that precession-driven LSCV  can sustain dynamos at relatively low magnetic Prandtl numbers ($P_m<1$). This allows us to investigate precession driven dynamos in the parameter regime in which the diffusivities have the correct hierarchy for planetary applications.        

The plan of this paper is as follows. Sec. \ref{sec:Model} introduces the governing equations and numerical models, while Sec. \ref{sec:Result} presents numerical results. The paper closes with a summary and discussion in Sec. \ref{sec:Diss}.

\section{Numerical models}\label{sec:Model}
\begin{figure}
\includegraphics[width=0.45\textwidth]{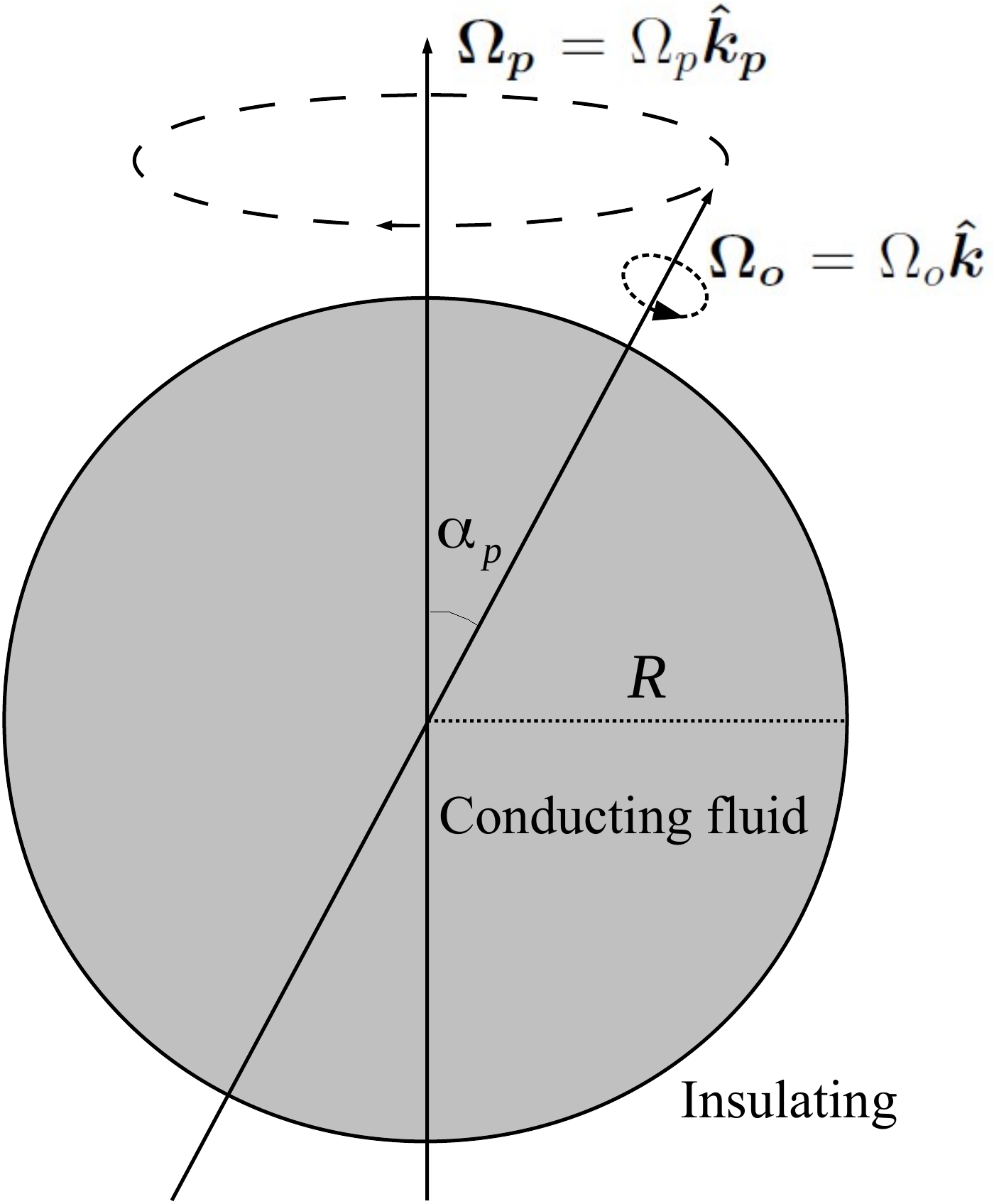}
\caption{Sketch of the problem. The precession axis $\bm{\Omega_p}$ is inclined at angle $\alpha_p$ to the rotation axis $\bm{\Omega_o}$.}
\label{fig6:problem}
\end{figure}
We consider a sphere of radius $R$ filled with a homogeneous, incompressible and electrically conducting fluid of density $\rho$, kinematic viscosity $\nu$, electrical conductivity $\sigma$ and magnetic permeability $\mu_0$ (equal to the vacuum magnetic permeability). The sphere rotates at $\bm{\Omega_o}=\Omega_o\bm{\hat{k}}$ and precesses at $\bm{\Omega_p}=\Omega_p\bm{\hat{k}_p}$, where $\bm{\hat{k}}$ and $\bm{\hat{k}_p}$ are unit vectors along the spin and precession axes, respectively (\autoref{fig6:problem}). Using the radius $R$ as the length scale, $\Omega_o^{-1}$ as the time scale and $\Omega_oR\sqrt{\rho \mu_0}$ as the unit of magnetic field $\bm B$, the dimensionless MHD equations governing the fluid velocity $\bm u$ and the magnetic field $\bm B$ in the mantle frame (attached to the container) can be written as, \cite{Tilgner2005, Zhang2010a}
  \begin{equation}\label{eq6:NS}
\frac{\partial \bm u}{\partial t}+\bm u\cdot \nabla \bm u+2(\bm{\hat{k}}+P_o \bm{\hat{k}_p})\times \bm u=-\nabla p+E\nabla^2\bm u-P_o(\bm{\hat{k}_p}\times \bm{\hat{k}})\times \bm r  +(\nabla\times \bm B)\times \bm B
\end{equation}
\begin{equation}\label{eq6:induction}
{ \frac{\partial \bm B}{\partial t}=\nabla \times (\bm u \times \bm B)+\frac{E}{P_m}\nabla^2\bm B}
\end{equation}
where the precession vector $\bm{\hat k_p}$ is given by
\begin{equation}
\bm{\hat k_p}=\sin(\alpha_p)\cos(t)\bm{\hat \imath}-\sin(\alpha_p)\sin(t)\bm{\hat \jmath}+\cos(\alpha_p)\bm{\hat k}.
\end{equation}
Here $(\bm{\hat \imath},\bm{\hat \jmath},\bm{\hat k})$ are unit vectors in  Cartesian coordinates $(x,y,z)$ whose $z$-axis is along the rotation vector $\bm{\hat k}$ and $\alpha_p$ is the angle between the rotation axis and the precession axis. We set $\alpha_p=60^\circ$ in all simulations unless otherwise specified. 

The system is controlled by three dimensionless parameters: the Ekman number $E$ which measures the ratio between the viscous force and the Coriolis force, the Poincar\'e number $P_o$ which measures the dimensionless precession rate and the magnetic Prandtl number $P_m$ which is the ratio between the kinematic viscosity and the magnetic diffusivity. These parameters are  defined as follows: 
$$
 E=\frac{\nu}{\Omega_o R^2}, \quad P_o=\frac{\Omega_p}{\Omega_o}, \quad  P_m=\nu\sigma \mu_0=\frac{\nu}{\eta},
$$
where $\eta=(\sigma \mu_0)^{-1}$ is the magnetic diffusivity of the fluid. Negative (positive) values of $P_o$ correspond to retrograde (prograde) precession. We consider only retrograde precession (negative $P_o$) in the present study.

Equations (\ref{eq6:NS}-\ref{eq6:induction}) are numerically solved by a fully spectral code. \cite{MartiThesis2012, Marti2016}  The velocity field $\bm u$ and magnetic field $\bm B$ are decomposed into toroidal and poloidal fields in a spherical coordinate system $(r,\theta,\phi)$:
\begin{equation}
\bm u=\nabla\times(T \bm{r})+\nabla\times\nabla\times(P \bm{r}),
\label{eq6:TPDu}
\end{equation}
\begin{equation}
\bm B=\nabla\times(\mathcal T \bm{r})+\nabla\times\nabla\times(\mathcal P \bm{r}),
\label{eq6:TPDB}
\end{equation}     
which automatically satisfy $\nabla\cdot \bm u=0$ and $\nabla\cdot \bm B=0$. The scalar fields are then expanded as
 \begin{equation}
 T(r,\theta,\phi)=\sum_{l=1}^L\sum_{m=-l^*}^{l^*} T_{l}^m(r)Y_l^m(\theta,\phi),
\label{eq6:TEx}
\end{equation}  
 \begin{equation}
 P(r,\theta,\phi)=\sum_{l=1}^L\sum_{m=-l^*}^{l^*} P_{l}^m(r) Y_l^m(\theta,\phi),
\label{eq6:PEx}
\end{equation} 
 \begin{equation}
 \mathcal T(r,\theta,\phi)=\sum_{l=1}^L\sum_{m=-l^*}^{l^*} \mathcal T_{l}^m(r)Y_l^m(\theta,\phi),
\label{eq6:TExB}
\end{equation}  
 \begin{equation}
 \mathcal P(r,\theta,\phi)=\sum_{l=1}^L\sum_{m=-l^*}^{l^*} \mathcal P_{l}^m(r) Y_l^m(\theta,\phi),
\label{eq6:PExB}
\end{equation} 
where $Y_l^m(\theta,\phi)$ are the spherical harmonics of degree $l$ and order $m$. Note that we may use $M<L$ as a restricted truncation in azimuth, and thus $l^*=${min}$(l,M)$. The radial dependences of the scalar fields are expanded in the so-called Worland polynomials $W_n^l(r)$, i.e.  $W_n^l(r)=r^l P_n^{-1/2,l-1/2}(2r^2-1)$, which are combinations of a prefactor $r^l$ and the one-sided Jacobi polynomials. The Worland polynomials exactly satisfy the parity and regularity at the origin of the sphere. \cite{Livermore2007} We use a total of $N$ polynomials for each $l$. Some of our more intensive calculations require truncations in ($N,L,M$) as high as (127,255,127).

The no-slip boundary condition for the velocity $\bm u$ is adopted and given by \cite{MartiThesis2012}
\begin{equation}
T_l^m(r)|_{r=1}=0,\quad P_l^m(r)|_{r=1}=0, \quad \frac{\partial }{\partial r} P_l^m(r)|_{r=1}=0.
\end{equation}

For the magnetic field $\bm B$, we use an insulating boundary condition which leads to a vanishing toroidal component and the poloidal component matching the potential magnetic field outside the sphere \cite{MartiThesis2012}
\begin{equation}
\mathcal T_l^m(r)=0,
\end{equation} 
\begin{equation}
\frac{\partial}{\partial r}\mathcal P_l^m(r)+\frac{l+1}{r} \mathcal P_l^m(r)=0,
\end{equation} 
on the boundary $r=1$. 
 
The numerical code has been benchmarked in several contexts including that of precession driven flows and MHD calculations. \cite{Hollerbach2013,Marti2014, Marti2016}

\section{Results}\label{sec:Result}
\renewcommand{\baselinestretch}{1.3}\normalsize
\begin{table}
\begin{center}
\begin{tabular}{cccccclcc}
\hline
$E$ & $P_o$ & $P_m$  & $(N,L,M)$  & Simulation time ($\tau$) & Dynamo & Figure  \\
\hline
$1.4\times 10^{-3}$ & -0.3 & 12  & (63,63,63) & 1.53  & yes & Fig. \ref{fig6:E_Pm} \\
$1.4\times 10^{-3}$ & -0.3 & 11  & (63,63,63)  & 0.46 & yes & Fig. \ref{fig6:E_Pm}   \\
$1.4\times 10^{-3}$ & -0.3 & 10  & (63,63,63)  & 0.9  & yes & Fig. \ref{fig6:E_Pm}  \\
$1.4\times 10^{-3}$ & -0.3 & 9 & (63,63,63)  & 0.35  & no  & Fig. \ref{fig6:E_Pm} \\
$1.4\times 10^{-3}$ & -0.3 & 8 & (63,63,63)  & 0.24  & no  & Fig. \ref{fig6:E_Pm} \\
$1.2\times 10^{-3}$ & -0.3 & 12  & (63,63,63)  & 0.2  & yes & Fig. \ref{fig6:E_Pm}  \\
$1.2\times 10^{-3}$ & -0.3 & 10  & (63,63,63)  & 0.42 & yes  & Fig. \ref{fig6:E_Pm}  \\
$1.2\times 10^{-3}$ & -0.3 & 9 & (63,63,63) &  0.63  & yes & Fig. \ref{fig6:E_Pm} \\
$1.2\times 10^{-3}$ & -0.3 & 8 & (63,63,63) &  0.38  & no  & Fig. \ref{fig6:E_Pm} \\
$1.0\times 10^{-3}$ & -0.25 & 12 & (63,63,63)  & 0.37  & yes & Fig. \ref{fig6:E_Pm}  \\
$1.0\times 10^{-3}$ & -0.25 & 10 & (63,63,63)  & 0.65  & yes & Fig. \ref{fig6:E_Pm}  \\
$1.0\times 10^{-3}$ & -0.25 & 9  & (63,63,63)  & 0.24  & no  & Fig. \ref{fig6:E_Pm} \\
$1.0\times 10^{-3}$ & -0.25 & 8  & (63,63,63)  & 0.3  & no   & Fig. \ref{fig6:E_Pm}\\
$7.0\times 10^{-4}$ & -0.2 & 10  & (63,127,127)  & 0.1  & yes & Fig. \ref{fig6:E_Pm}  \\
$7.0\times 10^{-4}$ & -0.2 & 8 & (63,127,127) &  0.05  & yes  & Fig. \ref{fig6:E_Pm}\\
$7.0\times 10^{-4}$ & -0.2 & 6 & (63,127,127) &  0.9  & yes  & Figs. \ref{fig6:E_Pm}, \ref{fig6:uBE7e-4Pm6},  \ref{fig6:Spectra_E7e-4Pm6} \\
$7.0\times 10^{-4}$ & -0.2 & 4 & (63,127,127) &  1.24  & yes  & Fig. \ref{fig6:E_Pm} \\
$7.0\times 10^{-4}$ & -0.2 & 2 & (63,127,127) &  0.22  & no  & Fig. \ref{fig6:E_Pm} \\
$5.0\times 10^{-4}$ & -0.08 & 12 & (63,127,127) & 0.1   & yes & Fig. \ref{fig6:E_Pm}  \\
$5.0\times 10^{-4}$ & -0.08 & 10 & (63,127,127) & 1.15  & yes & Fig. \ref{fig6:E_Pm}  \\
$5.0\times 10^{-4}$ & -0.08 & 8 & (63,127,127) &  0.46 & no  & Fig. \ref{fig6:E_Pm} \\
$5.0\times 10^{-4}$ & -0.08 & 6 & (63,127,127) &  0.17& no & Fig. \ref{fig6:E_Pm} \\
$3.0\times 10^{-4}$ & -0.05 & 12 & (63,127,127) & 0.1 & no & Fig. \ref{fig6:E_Pm}  \\
$3.0\times 10^{-4}$ & -0.05 & 10 & (63,127,127) & 0.05   & no & Fig. \ref{fig6:E_Pm} \\
$1.0\times 10^{-4}$ & -0.1 ($\alpha_p=90^{\circ}$) & 1  & (127,127,127) & 0.15 &  yes  & Fig. \ref{fig6:spectra_E1e-4Pm1} \\
$6.0\times 10^{-5}$ & $-2\times 10^{-2}$ & 2 & (127,255,127)  & 0.8   & yes & Fig. \ref{fig6:E6e-5Pm2}  \\
$6.0\times 10^{-5}$ & $-2\times 10^{-2}$ & 0.5 & (63,127,127)  & 0.56   & no & - \\
$3.0\times 10^{-5}$ & $-1.35\times 10^{-2}$ & 0.5 & (127,255,127)  & 1.2   & yes & Figs. \ref{fig6:energy_E3e-5Pm05}-\ref{fig6:spectra_E3e-5Pm05}  \\

\hline
\end{tabular}
\end{center}
\caption{Parameters of all simulation in this study. $E$, $P_o$, $P_m$ are the Ekman number, Poincar\'e number and magnetic Prandtl number respectively. $N, L, M$ are truncations in radius, latitude, and azimuth. $\tau=\frac{P_m}{E}\Omega_o^{-1}$ is the magnetic diffusion time. }
\label{tab:models}
\end{table}

\renewcommand{\baselinestretch}{1.5}
The simulations performed for this work are detailed in  \autoref{tab:models}. We use Ekman numbers in the range $3.0\times 10^{-5}\leqslant E \leqslant 1.4 \times 10^{-3}$, magnetic Prandtl number $0.5 \leqslant P_m \leqslant 12$, and Poincar\'e number  $1.35\times 10^{-2}  \leqslant -P_o \leqslant 0.3$. 
The numerical simulations are diagnosed by the total kinetic energy $E_k$ in the fluid volume
\begin{equation}
E_k=\frac{1}{2}\int |\bm u|^2 \mathrm{d} V,
\end{equation}
and the magnetic energy $E_m$ in the fluid volume 
\begin{equation}
E_m=\frac{1}{2}\int |\bm B|^2 \mathrm{d} V.
\end{equation}
The kinetic energy can be decomposed into its symmetric part and anti-symmetric part as we have done in hydrodynamic simulations. \cite{Lin2015} 
\begin{equation}
E_{ks}=\frac{1}{2}\int |\bm u_s|^2 \mathrm{d} V,\quad E_{ka}=\frac{1}{2}\int |\bm u_a|^2 \mathrm{d} V ,
\end{equation}
where 
\begin{equation}
\bm u_s=\frac{\bm u(\bm r)-\bm u(-\bm r)}{2}, \quad \bm u_a=\frac{\bm u(\bm r)+\bm u(-\bm r)}{2}.
\end{equation}
The basic flow driven by precession is symmetric around the origin and any anti-symmetric flows must be due to instabilities. Therefore, the anti-symmetric kinetic energy is an indicator of instabilities.

All results are presented in the mantle frame in order to have the same view as that from which we observe the Earth's magnetic field.

\subsection{Laminar dynamos}
It has been shown that precession driven laminar flows can sustain dynamos due to the Ekman pumping/suction at large Ekman numbers. \cite{Tilgner2005} In this section, we gradually reduce the Ekman number and the Poincar\'e number to see whether precession driven laminar flows can sustain dynamos at lower Ekman numbers $E$ and if so what would be the critical magnetic Prandtl number $P_m$.

\begin{figure}
\begin{center}
\includegraphics[width=0.49\textwidth]{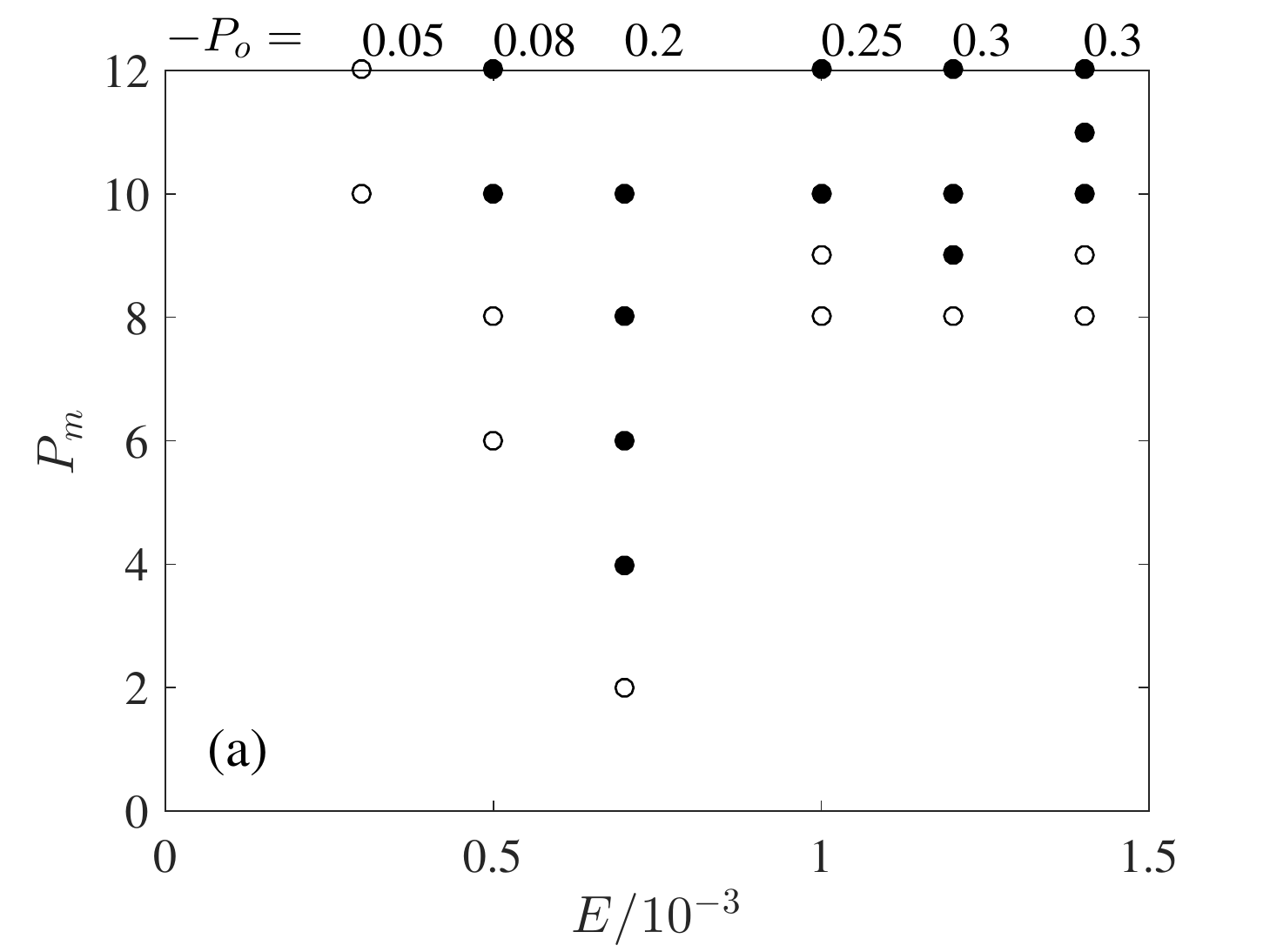}
\includegraphics[width=0.49\textwidth]{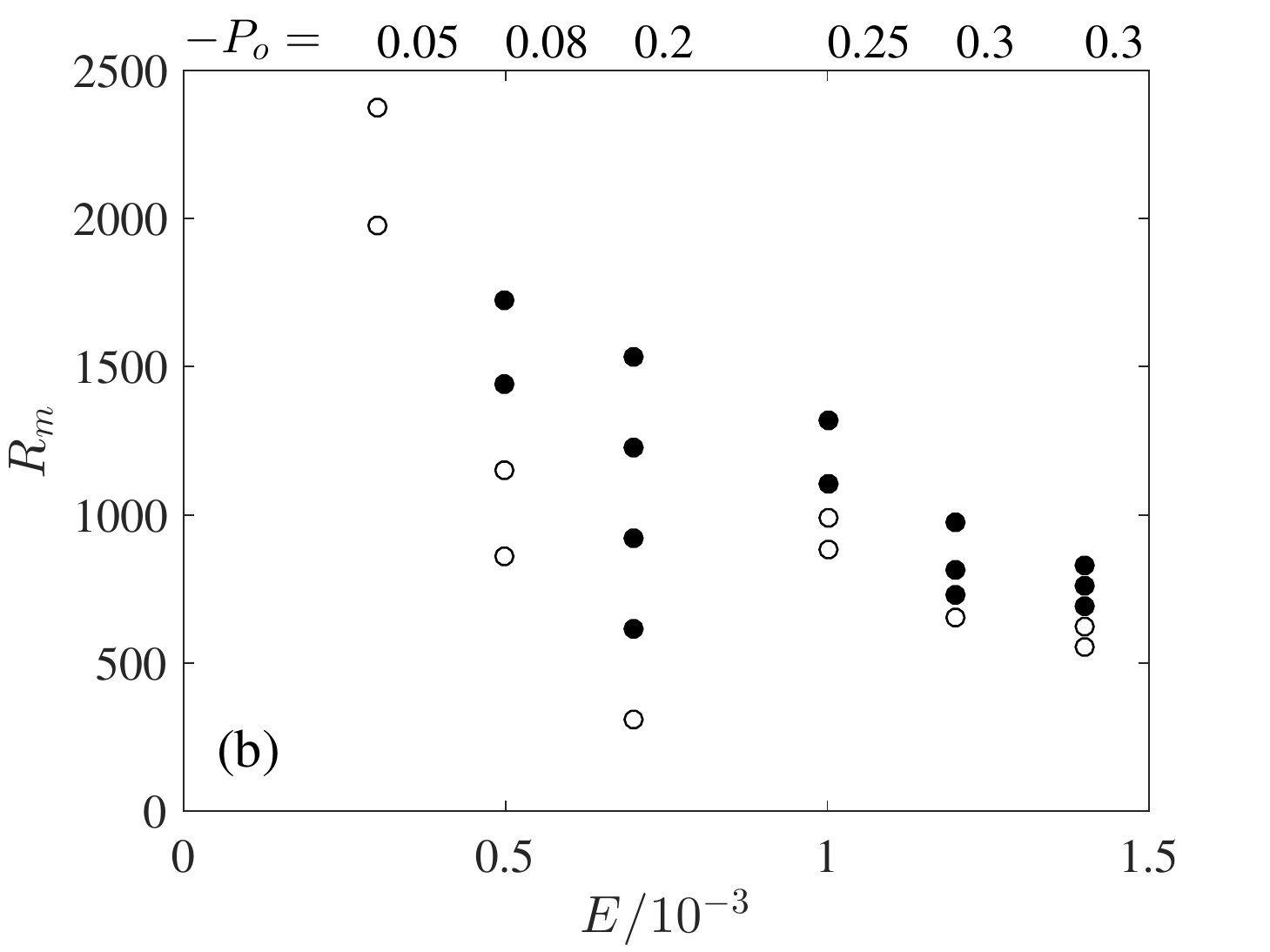}
\caption{Regime diagram of the laminar dynamo simulations in the parameter space of (a) magnetic Prandtl numbers and Ekman numbers ($P_m$, $E$) and (b) magnetic Reynolds numbers and Ekman numbers ($R_m$, $E$). At each given Ekman number, the corresponding Poincar\'e number is given at the top of plots. It was chosen such that the flow is in the hydrodynamically stable regime. Filled circles represent dynamo action and open circles represent failed dynamos.} 
\label{fig6:E_Pm}
\end{center}
\end{figure}

\begin{figure}
\begin{center}
\includegraphics[width=0.49 \textwidth,clip,trim=2cm 2cm 0 0]{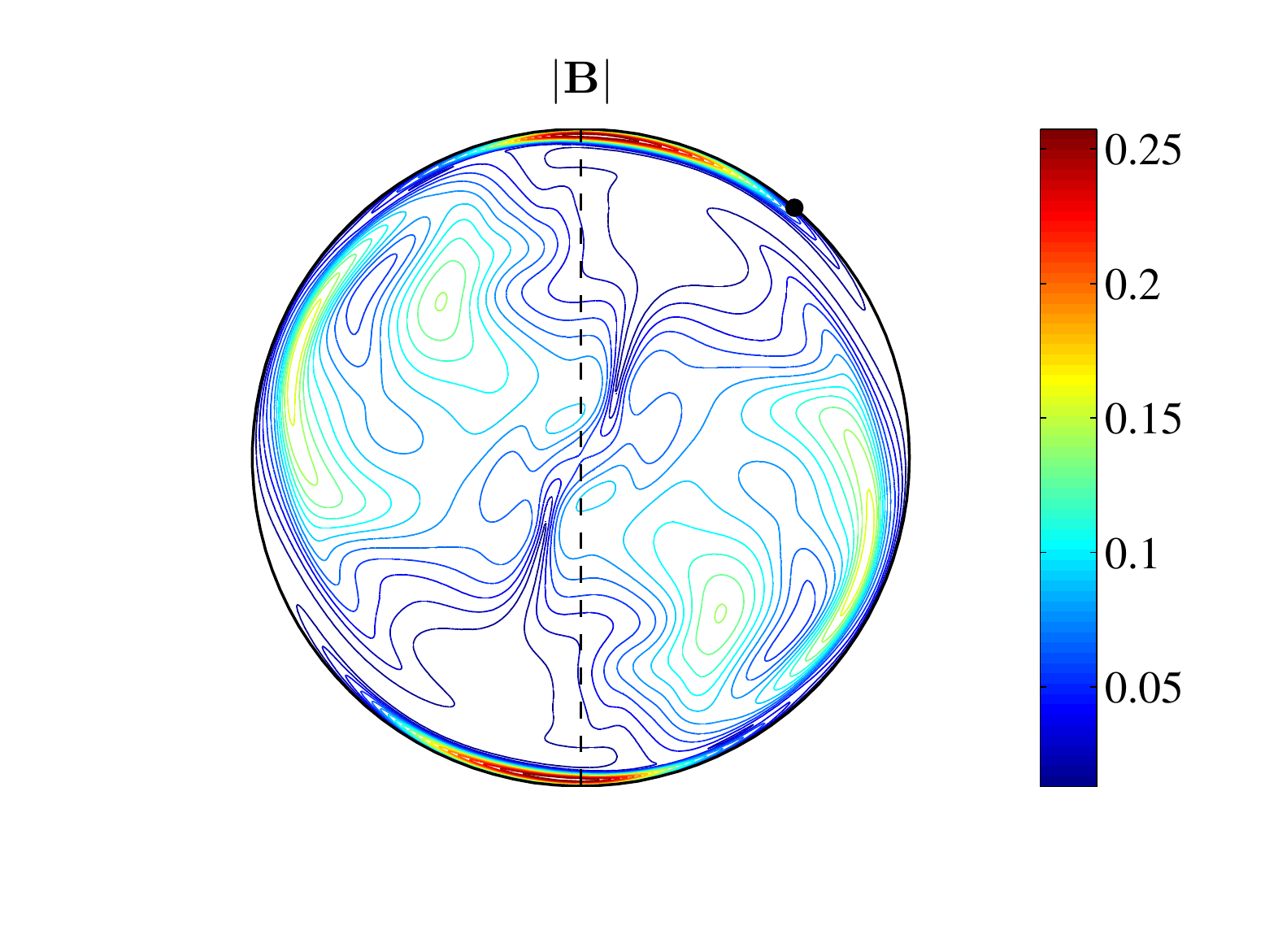}
\includegraphics[width=0.49 \textwidth,clip,trim=2cm 2cm 0 0]{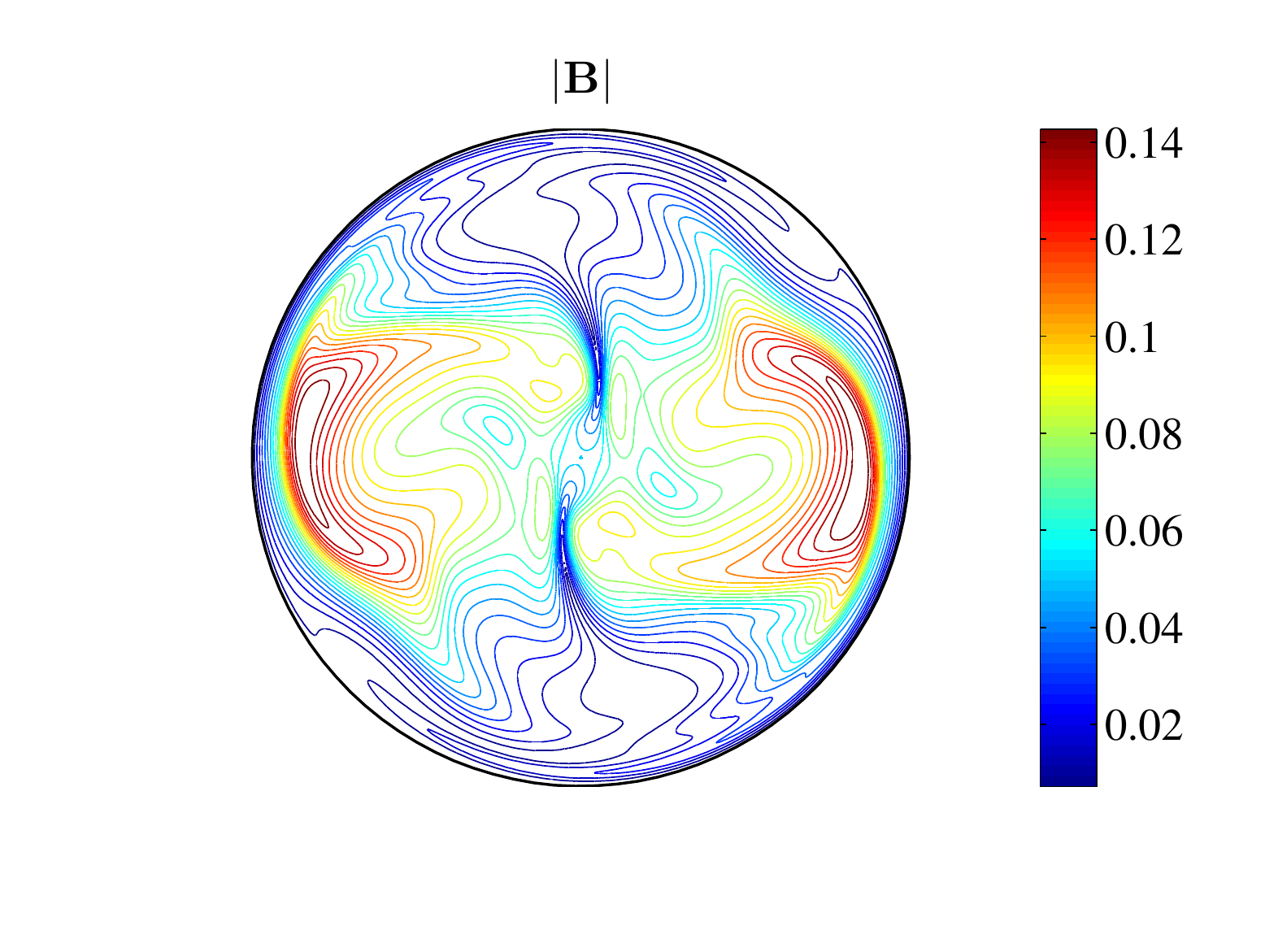} \\
(a) \hspace*{7cm} (b) \\
\includegraphics[width=0.49 \textwidth,clip,trim=0cm 0cm 0 0cm]{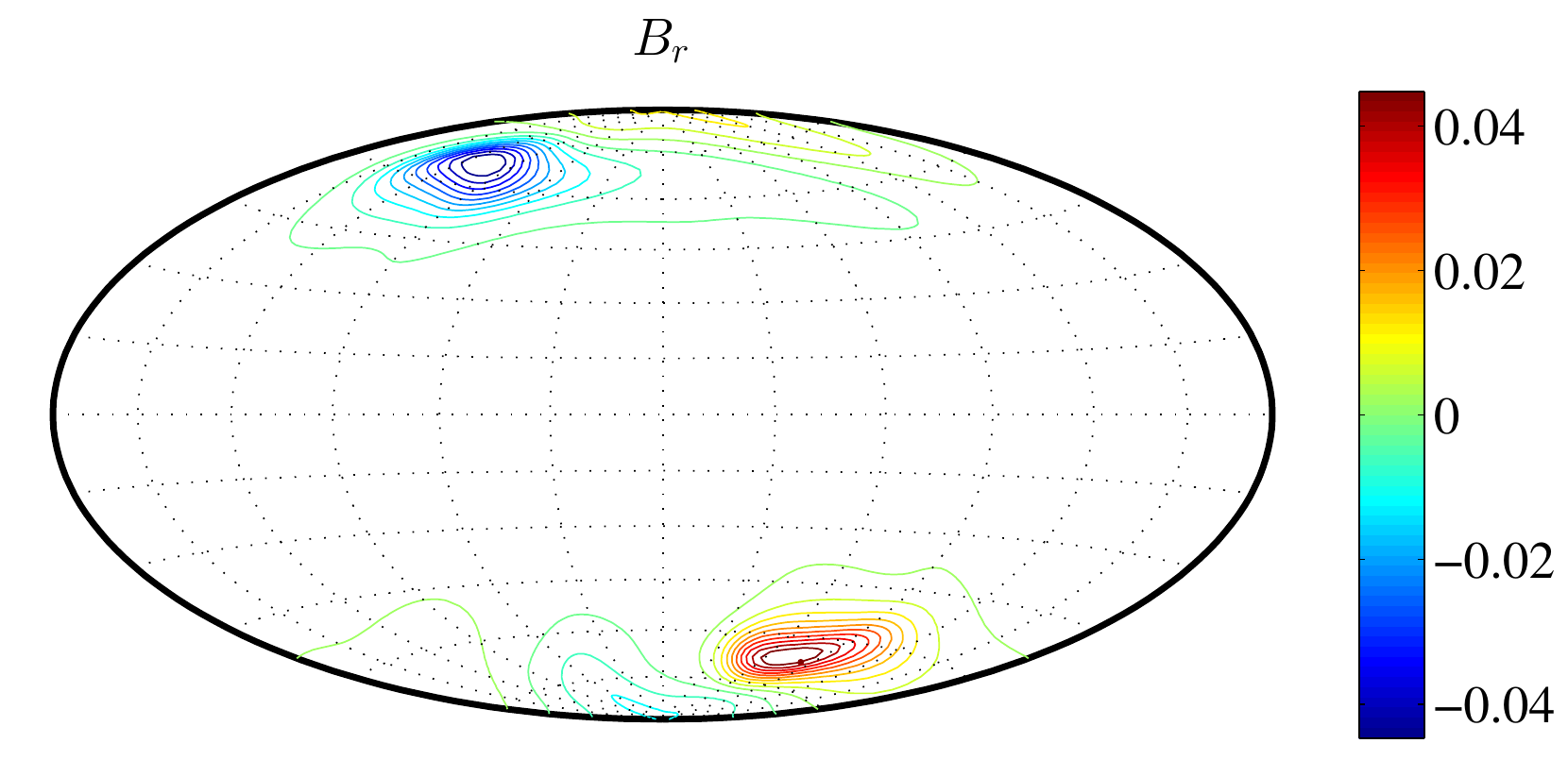}
\includegraphics[width=0.49 \textwidth,clip,trim=0cm 0cm 0 0cm]{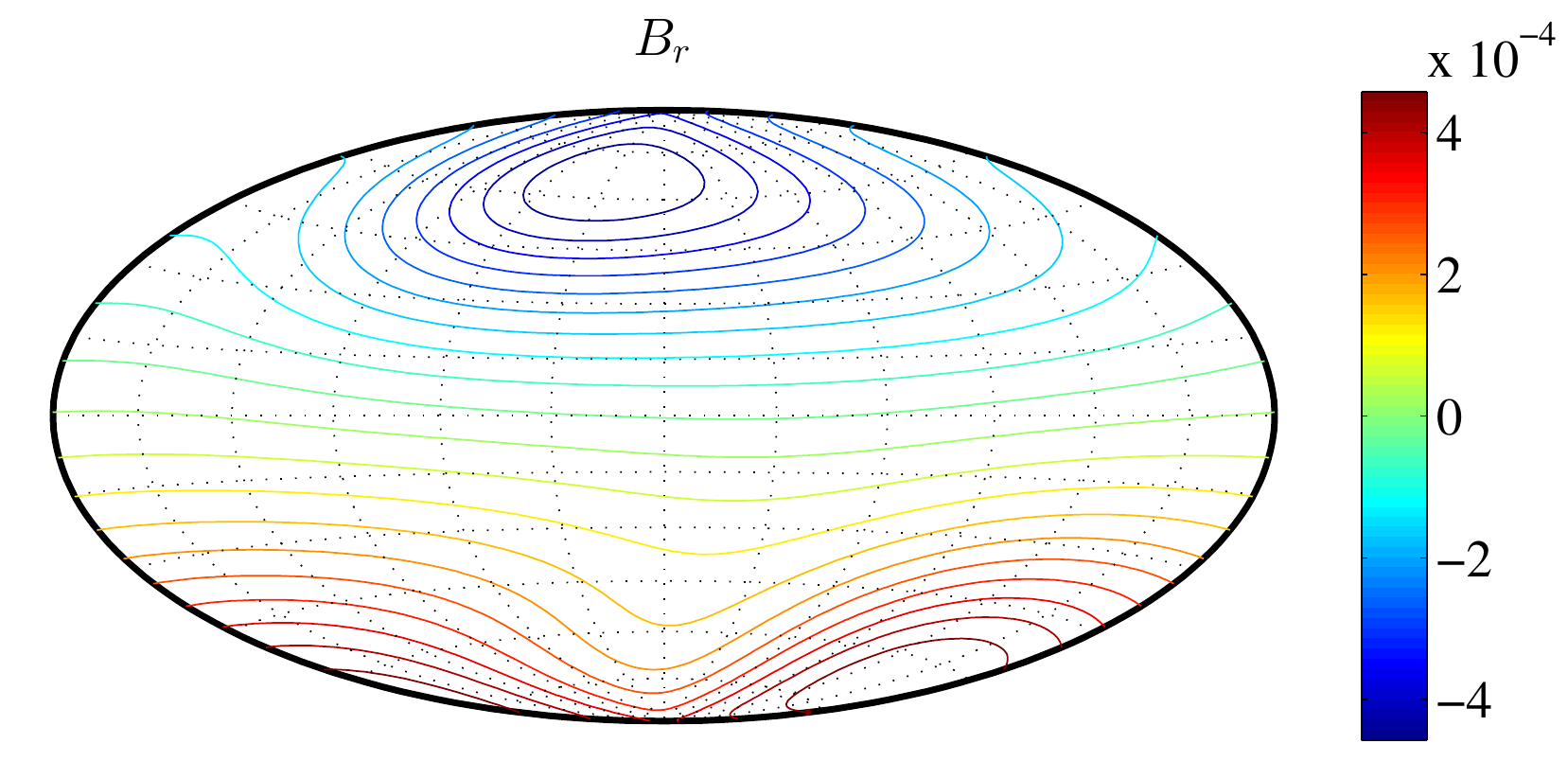} \\
(c) \hspace*{7cm} (d)
\caption{Snapshot of the magnetic field generated by laminar flow at $E=7\times 10^{-4}$, $P_o=-0.2$, $P_m=6$. (a)  Magnetic field strength $|\bm B|$ in a meridional plane. The dashed line is the rotation axis of the container.  Black dots represent the position of the rotation axis of the fluid. (b) Magnetic field strength $|\bm B|$ in the equatorial plane with respect to the rotation axis of the container. (c) Radial component of the magnetic field $B_r$ on the surface of $r=1-\sqrt{E}$ (Hammer projection). (d) Radial component of the magnetic field $B_r$ on the surface of $r=2$ (Hammer projection).} 
\label{fig6:uBE7e-4Pm6}
\end{center}
\end{figure}
\begin{figure}
\begin{center}
\includegraphics[width=0.98\textwidth]{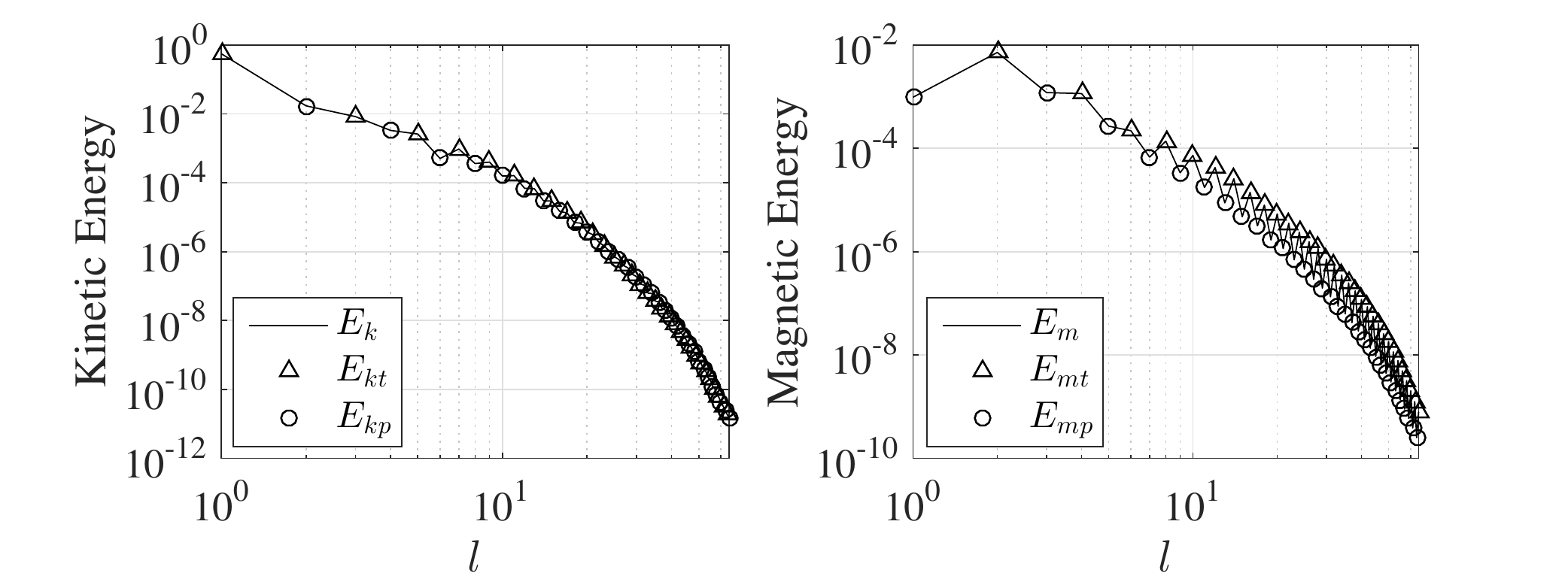} \\
(a)\hspace*{6cm}(b)
\caption{Energy spectra of the laminar dynamo in \autoref{fig6:uBE7e-4Pm6} at $E=7\times 10^{-4}$, $P_o=-0.2$, $P_m=6$. (a) Kinetic energy spectra as a function of spherical harmonic degree $l$. (b) Magnetic energy spectra as a function of spherical harmonic degree $l$. Subscripts $t$ and $p$ represent the toroidal component and the poloidal component respectively.}
\label{fig6:Spectra_E7e-4Pm6}
\end{center}
\end{figure}
\begin{figure}
\begin{center}
\includegraphics[width=0.65\textwidth]{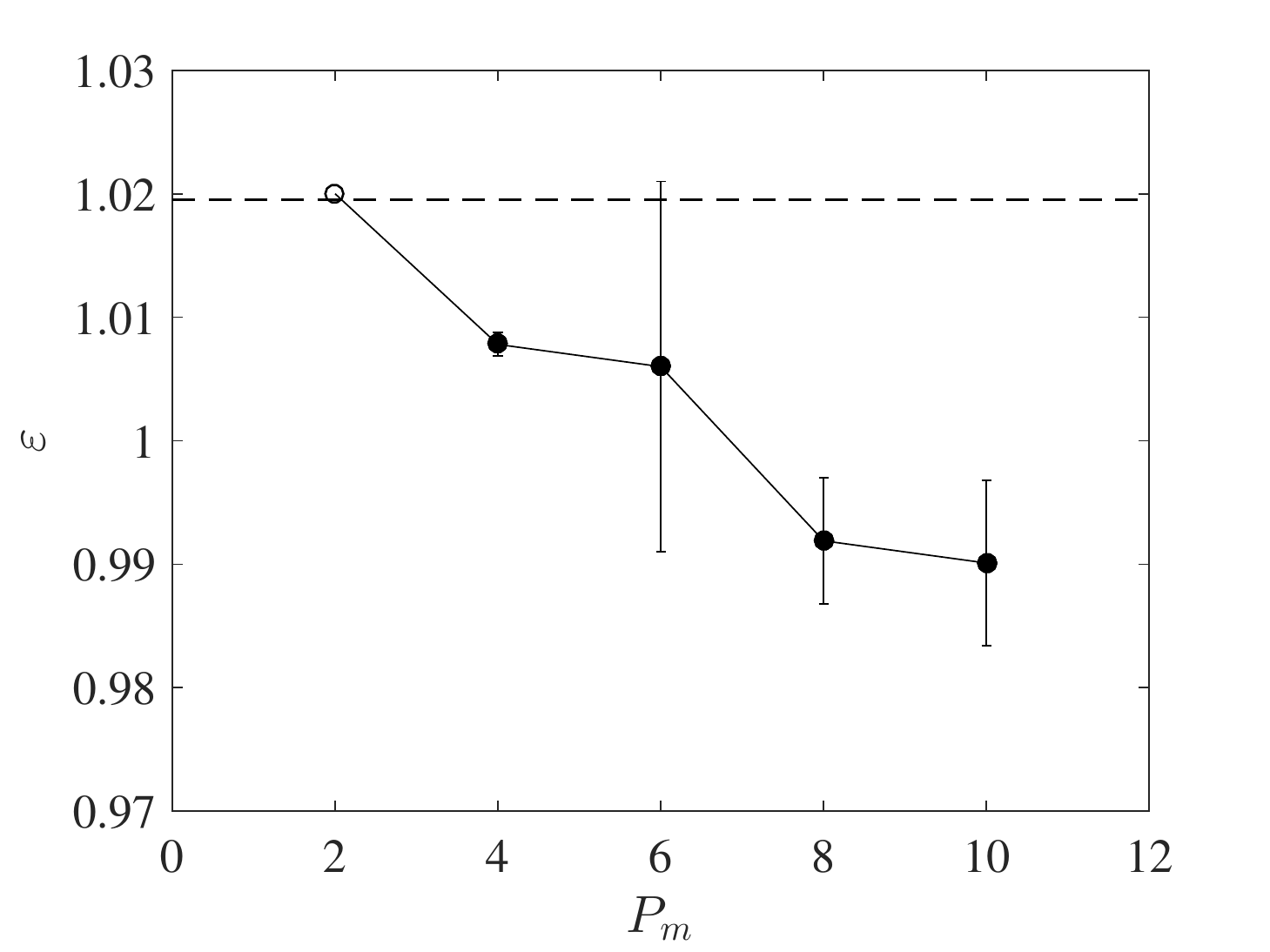} \\
\caption{The differential rotation $\varepsilon$ between the container and the fluid in the bulk as a function of $P_m$ at $E=7\times 10^{-4}$ and $P_o=-0.2$. The horizontal dashed line is the value of purely hydrodynamic simulation. The open circle (at $P_m=2$) represents no dynamo action. Filled circles (dynamo action) show the mean value after the saturation and the error bars show the standard deviation from the mean value.}
\label{fig6:PmDf}
\end{center}
\end{figure}

At a given Ekman number, we choose a suitable Poincar\'e number such that the flow is in the stable regime while containing as much kinetic energy as possible based on the hydrodynamic simulations (see Fig. 11 in Ref \onlinecite{Lin2015}). For each combination of $E$ and $P_o$, we vary the magnetic Prandtl number $P_m$ in order to find a critical value. The velocity field starts from a steady state of the hydrodynamic simulation while the magnetic field starts from a saturated dynamo state at higher $E$ or $P_m$.  If the magnetic energy can be sustained for around one magnetic diffusion time $\tau$:
\begin{equation}
\tau=\dfrac{R^2}{\eta}=\dfrac{\Omega_o R^2}{\eta}\Omega_o^{-1}=\dfrac{P_m}{E}\Omega_o^{-1} 
\end{equation}
then we say it is a successful dynamo. Note that $\tau$ is around 10 times the $e-$folding time of the slowest decaying dipole field. \cite{Roberts2007} Failed dynamos show exponential decay of the magnetic energy.  
\autoref{fig6:E_Pm} (a) shows the results in the plane ($P_m$, $E$), in which successful dynamos are marked as filled circles and failed ones are marked as open circles. We can see that the critical $P_m$ is around 10 for $E\geqslant 10^{-3}$ which 
then drops to 3 at $E=7\times 10^{-4}$. This sharp jump in the regime diagram may be attributed to the fact that Ekman numbers $E\geqslant 10^{-3}$ are not sufficiently small to show asymptotic behaviour. A similar jump in the hydrodynamic instability diagram was also observed in our hydrodynamic simulations (see Fig. 11 in Ref \onlinecite{Lin2015}). It seems that the critical magnetic Prandtl number increases as we reduce the Ekman number ($E\leqslant 7\times 10^{-4}$).  In the poineering study by Tilgner\cite{Tilgner2005} on precession driven dynamos, he defined a magnetic Reynolds number for the laminar dynamo based on the characteristic poloidal velocity:
\begin{equation}
R_m=(2E_{kp}/V)^{1/2}P_m/E,
\end{equation}
 where $E_{kp}$ is the poloidal kinetic energy and $V$ is the volume of the sphere. \autoref{fig6:E_Pm} (b) shows the regime diagram in the plane of ($R_m$, $E$). The critical magnetic Reynolds number is around 700 for $E > 10^{-3}$, which is very close to the critical value 770 obtained by Tilgner \cite{Tilgner2005} at $E=1.4\times 10^{-3}$ (there is a small solid inner core with radius of $0.1R$ in his study). At smaller Ekman numbers ($E\leqslant 7\times 10^{-4}$), it seems that the critical magnetic Reynolds number increased as the Ekman number is decreased, at least for the $R_m$ defined above.
One should bear in mind that we also adjust the Poincar\'e number as the Ekman number is decreased in order to keep the flow laminar in \autoref{fig6:E_Pm}. The flow would become unstable if we decrease the Ekman number but fix the Poincar\'e number. In the unstable regime, the critical magnetic Prandtl number can be smaller than in the laminar regime. \cite{Tilgner2005} We will focus on dynamos driven by unstable flows in Section \ref{subs: lscv}.

 \autoref{fig6:uBE7e-4Pm6} shows an example of the magnetic field generated by the laminar flow at  $E=7\times 10^{-4}$, $P_o=-0.2$ and $P_m=6$.  The magnetic field strength $|\bm B|$ is plotted in a meridional plane across both the rotation axes of the container and the fluid (a), and in the equatorial plane with respect to the rotation axis of the container (b). We can see that the magnetic field is mainly generated next to the boundary due to the Ekman pumping. In the equatorial plane, we see also some contributions in the bulk fluid, which is likely related to the conical shear layers spawned from the critical latitudes. \cite{Kerswell1995, Hollerbach1995,Noir2001, Noir2001b, Kida2011} \autoref{fig6:uBE7e-4Pm6} (c-d) show the radial component of the magnetic field on the surface just below the boundary ($r=1-\sqrt{E}$) and on the surface $r=2$ (roughly corresponding to the Earth's surface if we assume the boundary at $r=1$ represents the core-mantle boundary). The magnetic field outside the fluid domain ($r>1$) is extended upwards as a potential field. We can clearly see a dipole magnetic field. However, the orientation of the dipole axis undergoes diurnal variation due to the variation of the rotation axis of the fluid in the mantle frame (see Movie 1 in supplemental material \cite{Note1}).   
 
 In \autoref{fig6:Spectra_E7e-4Pm6}, we show the kinetic energy (a) and the magnetic energy (b) spectra of the laminar dynamo in the fluid volume at $E=7\times 10^{-4}$, $P_o=-0.2$ and $P_m=6$. The kinetic energy is dominated by the spherical harmonic degree $l=1$ component, whereas the magnetic energy is dominated by $l=2$. Since the laminar flow driven by precession is symmetric around the origin, the kinetic energy is composed of odd degrees $l$ in the toroidal field and even $l$ in the poloidal field. In contrast, the magnetic energy is composed of even $l$ in the toroidal field and odd $l$ in the poloidal field, which means the magnetic field is antisymmetric around the origin, i.e. $\bm{B}(-\bm r)=\bm{B}(\bm r)$.   

It is of interest to examine the feedback of the Lorentz force on the fluid flow in laminar dynamos. The laminar flow is dominated by a solid body rotation in the bulk of the fluid $\bm{\omega_F}$, which can be extracted from the mean vorticity in the bulk. \cite{Lin2015} A simple measure of the magnetic effect on the laminar flow would be the change of $\bm{\omega_F}$ in MHD simulations with respect to the corresponding hydrodynamic one. In \autoref{fig6:PmDf}, we plot the differential rotation between the container and the fluid, $\varepsilon=|\bm{\hat{k}}-\bm{\omega_F}|$, as a function of $P_m$ at $E=7\times 10^{-4}$ and $P_o=-0.2$, compared to the hydrodynamic value. In the hydrodynamic simulation (horizontal dashed line) and also the MHD simulations with no dynamo action (open circle), the differential rotation $\varepsilon$ is steady after a transient stage. However, the differential rotation $\varepsilon$ starts to fluctuate as the magnetic field grows in simulations with the dynamo action (filled circles). In \autoref{fig6:PmDf}, black dots show the mean value of $\varepsilon$ after the saturation of the magnetic energy and the error bars show the standard deviation from the mean value. We can see that the mean value of $\varepsilon$ after the saturation slightly drops (around 3\% at $P_m=10$) with respect to the hydrodynamic value due to the action of the Lorentz force on the fluid. 
 
\subsection{Dynamos driven by large scale cyclonic vortices} \label{subs: lscv}
\begin{figure}
\includegraphics[width=0.7\textwidth]{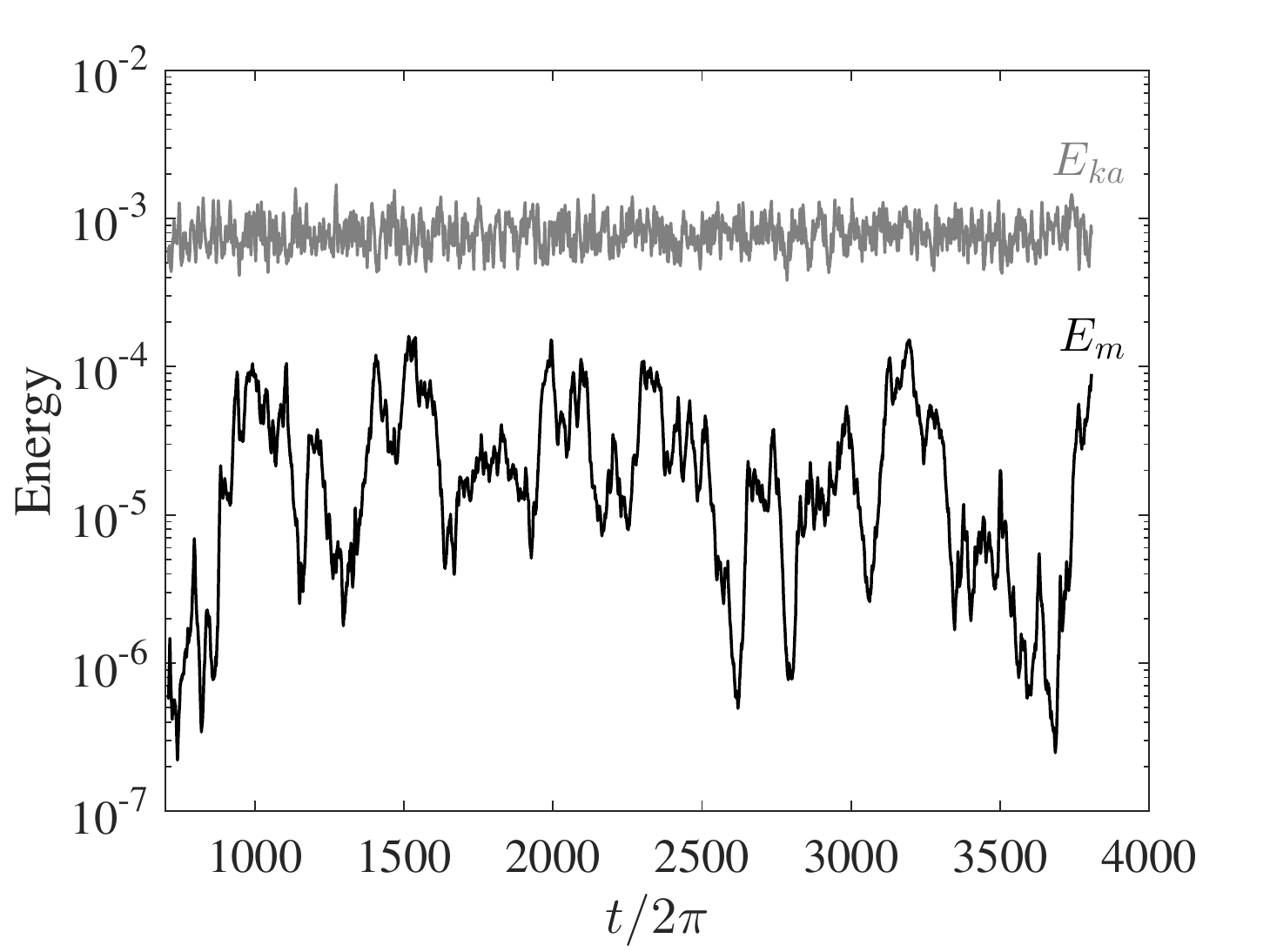}
\caption{Time evolution of the magnetic energy $E_m$ (black) and the antisymmetric kinetic energy $E_{ka}$ (grey) for parameters $E=3\times 10^{-5}$, $P_o=-1.35\times 10^{-2}$,  $P_m=0.5$.}
\label{fig6:energy_E3e-5Pm05}
\end{figure}

\begin{figure}
\includegraphics[width=0.9\textwidth,clip,trim=0 7cm 0 2cm]{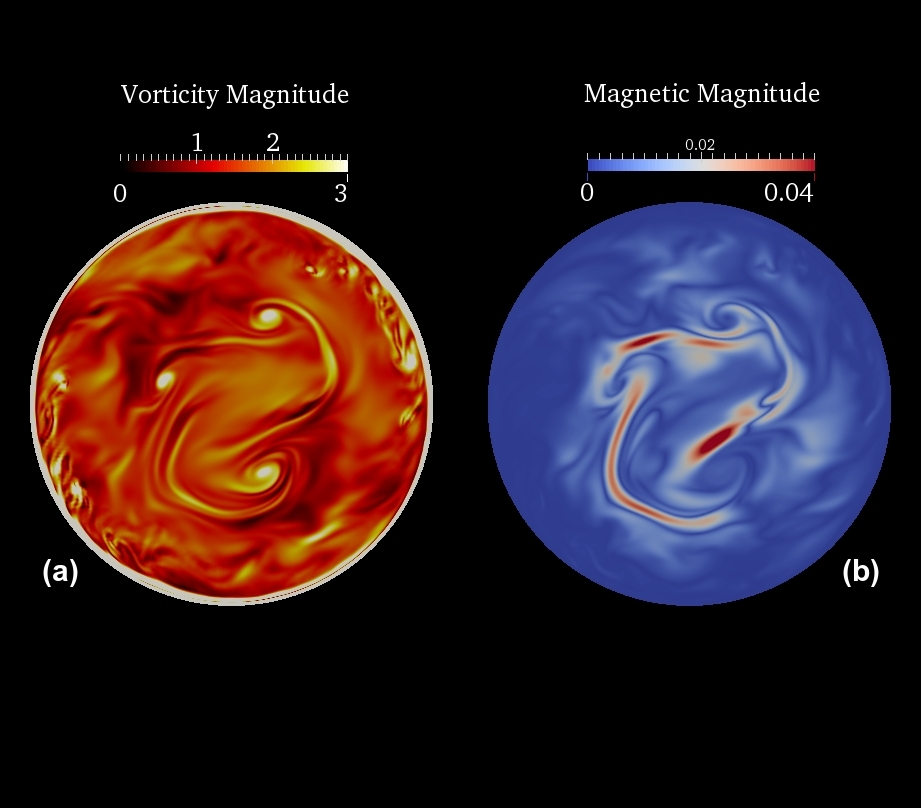}\\
\includegraphics[width=0.9\textwidth,clip,trim=0 7cm 0 2cm]{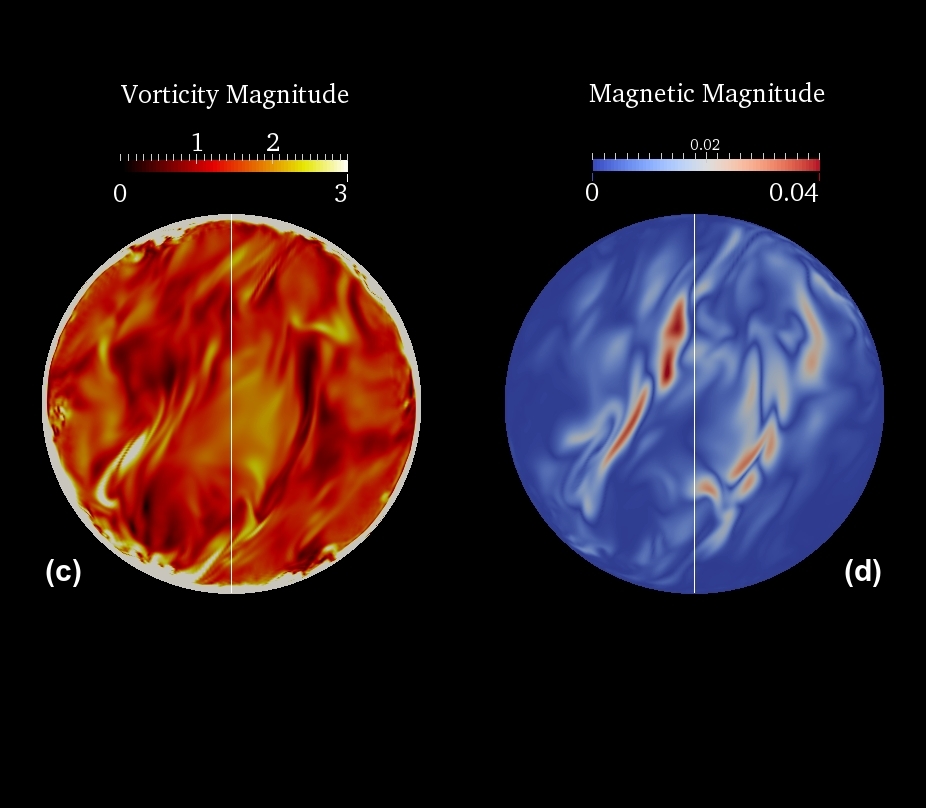}
\caption{Snapshot of the vorticity and the magnetic field at $t/2\pi=1505$, $E=3\times 10^{-5}$, $P_o=-1.35\times 10^{-2}$, $P_m=0.5$. (a) Vorticity $|\nabla \times \bm u|$ and (b) magnetic field strength $|\bm B|$ in the equatorial plane. (c) Vorticity and (d) magnetic field strength in a meridional plane. The white lines represent the rotation axis of the container.}
\label{fig6:visualization_E3e-5Pm05}
\end{figure}

\begin{figure}
\includegraphics[width=0.9\textwidth]{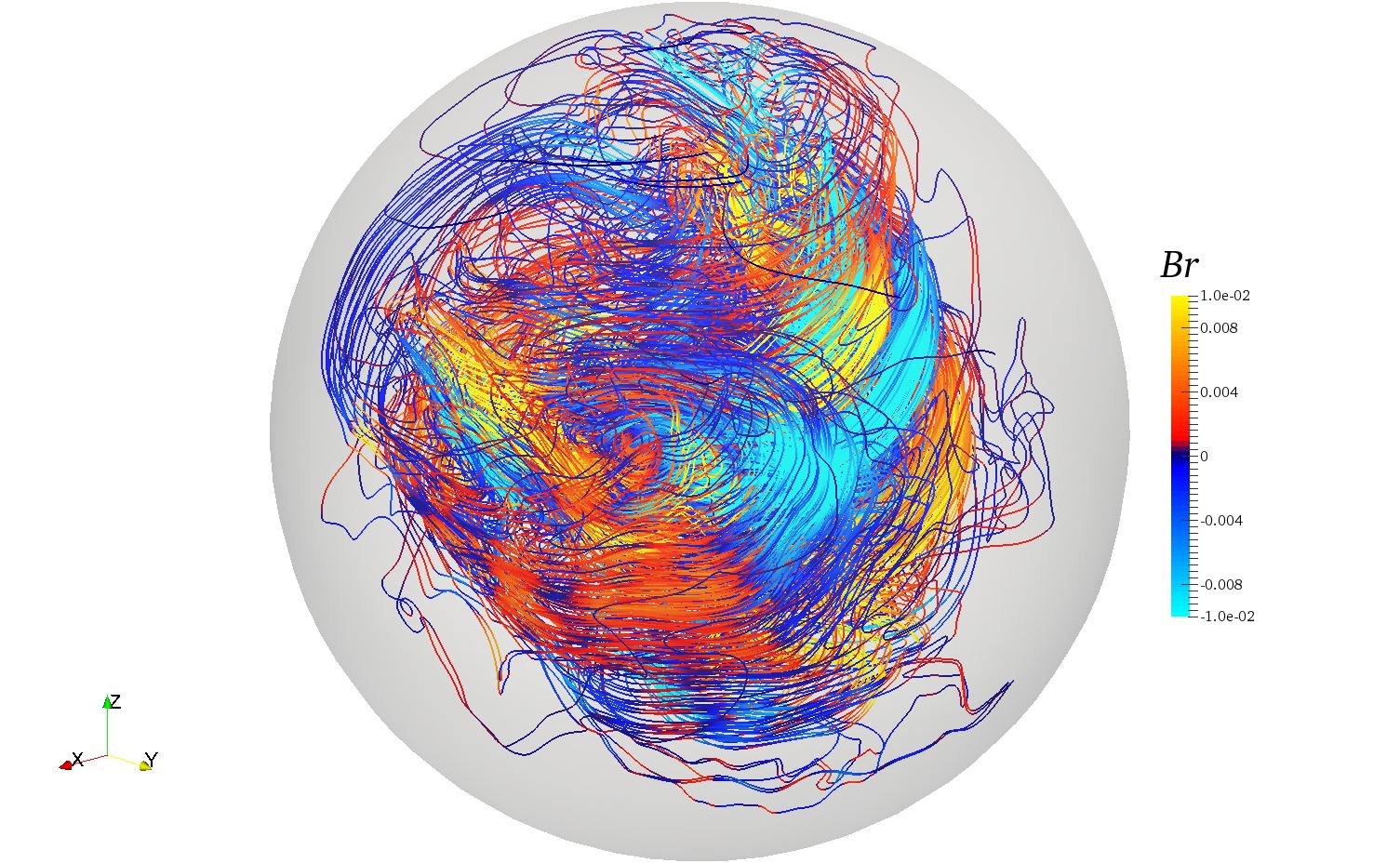}\\
\caption{A snapshot of the magnetic field lines colored by radial component of the magnetic field $B_r$ in the fluid domain at $t/2\pi=1509.4$,  $E=3\times 10^{-5}$, $P_o=-1.35\times 10^{-2}$, $P_m=0.5$. The field lines start from 100 points that are uniformly distributed on the spherical surface $r=0.5$ and have a maximum length of 20.}
\label{fig6:MagLines}
\end{figure}

\begin{figure}
 \includegraphics[width=0.49\textwidth]{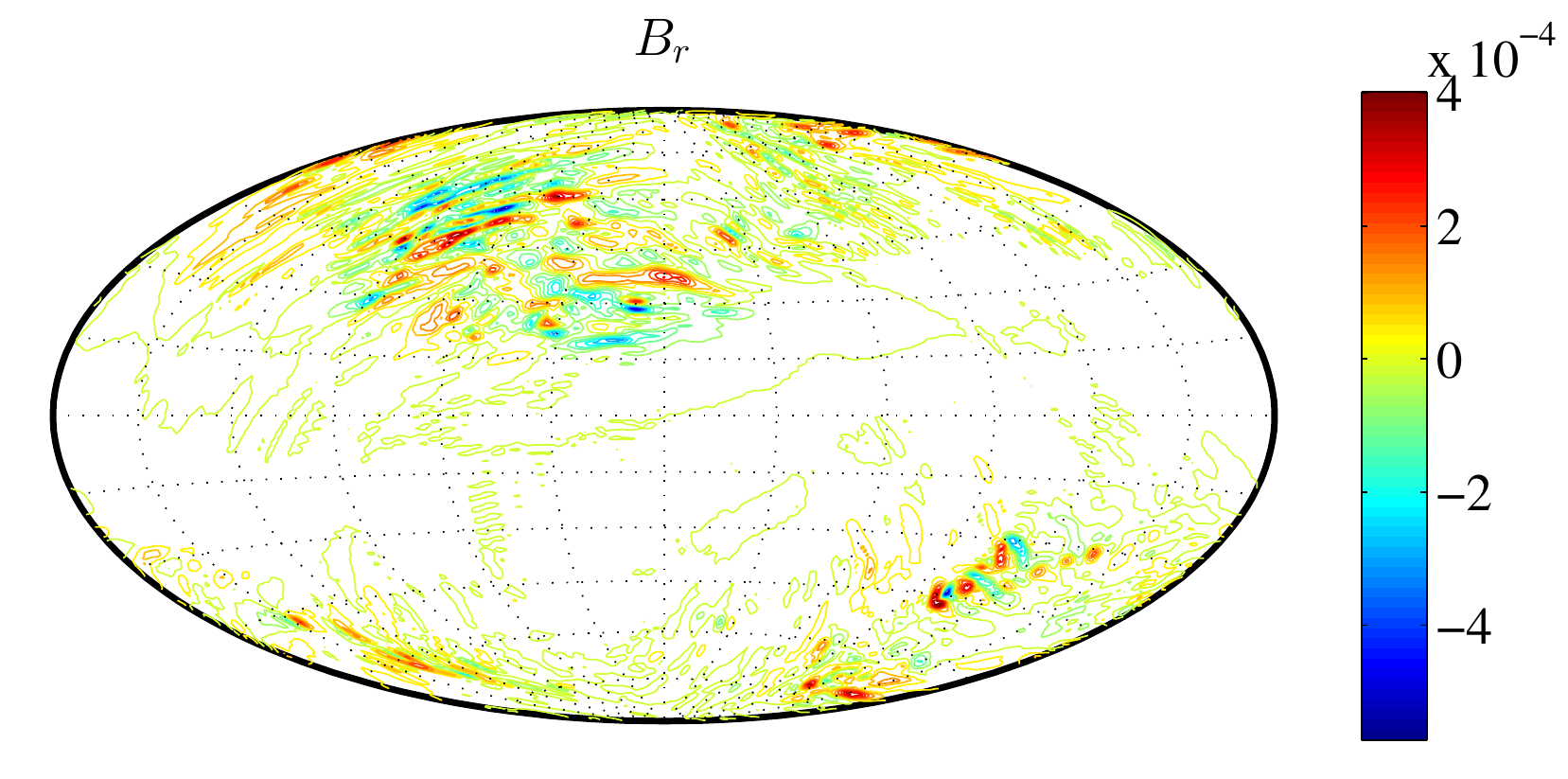}
 \includegraphics[width=0.49\textwidth]{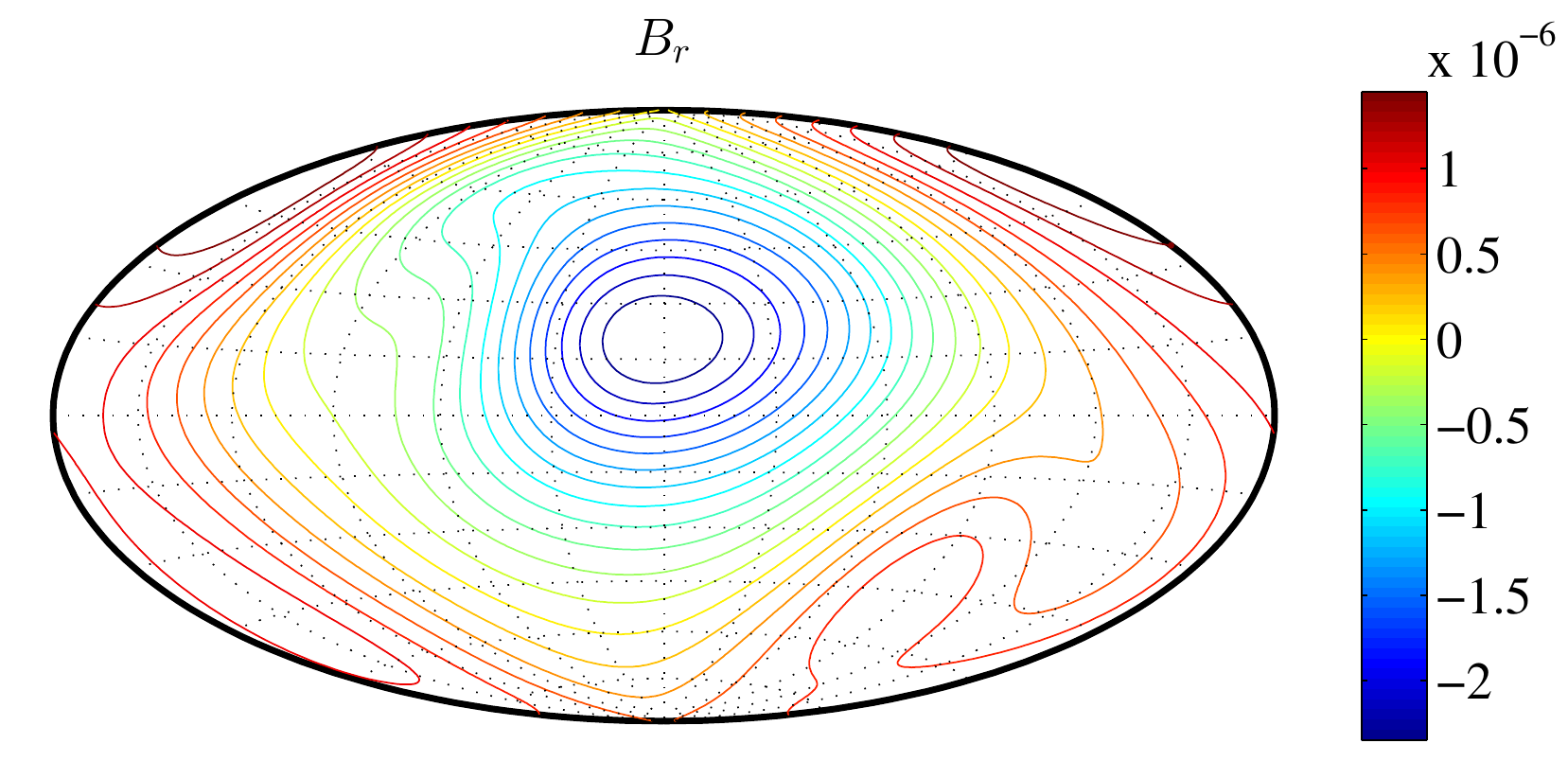}\\
 (a)\hspace*{6cm}(b)
 \caption{Radial component of the magnetic field $B_r$ at (a) the surface of $r=1-\sqrt{E}$  and (b) the surface of $r=2$. $t/2\pi=1505$, $E=3\times 10^{-5}$, $P_o=-1.35\times 10^{-2}$, $P_m=0.5$.}
 \label{fig6:BrBL}
\end{figure}

\begin{figure}
\includegraphics[width=0.95\textwidth]{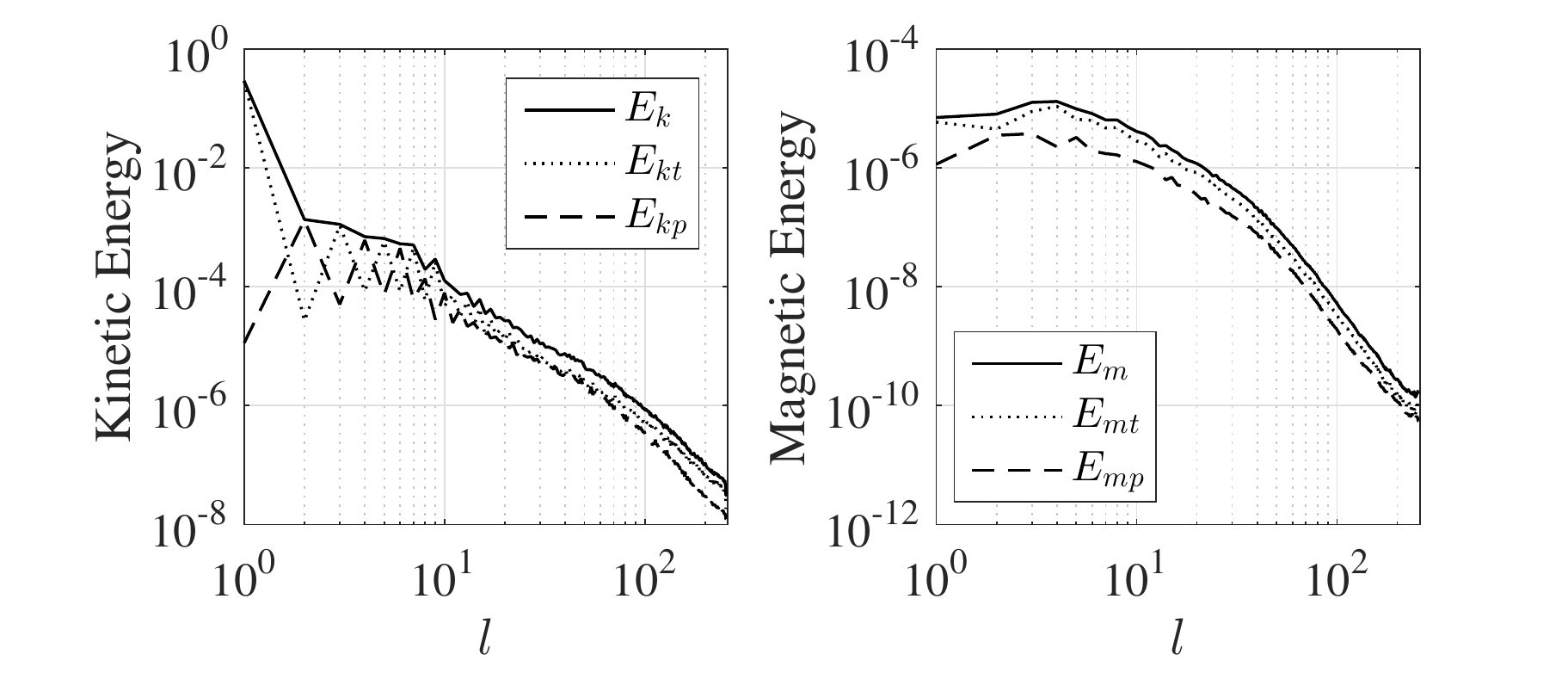} \\
(a)\hspace*{6cm}(b)
\caption{Snapshot of the energy spectra at $t/2\pi=1505$, $E=3\times 10^{-5}$, $P_o=-1.35\times 10^{-2}$, $P_m=0.5$. (a) Kinetic energy spectra versus spherical harmonic degree $l$. (b) Magnetic energy spectra versus spherical harmonic degree $l$. Subscripts $t$ and $p$ represent the toroidal component and the poloidal component respectively.}
\label{fig6:spectra_E3e-5Pm05}
\end{figure}

Both laboratory experiment \cite{Mouhali2012} and numerical simulations \cite{Lin2015} have shown that the nonlinear evolution of precessional instabilities may lead to large scale cyclonic vortices. Recent simulations have shown the robustness of the dynamo driven by large scale vortices in rotating convection. \cite{Guervilly2015} In this section, we focus on the magnetic field generation by large scale cyclonic vortices driven by precession.

We consider a case of $E=3\times 10^{-5}$ and $P_o=-1.35\times 10^{-2}$, at which the hydrodynamic simulation has shown the formation of the large scale cyclonic vortices (Fig. 10 in Ref \onlinecite{Lin2015}). \autoref{fig6:energy_E3e-5Pm05} shows the energy evolution of the MHD simulation at $P_m=0.5$. The velocity field starts from a saturated state in the hydrodynamic simulation and the magnetic field starts from small random perturbations. 
The total kinetic energy is dominated by the solid body rotation component so we plot the antisymmetric kinetic energy $E_{ka}$ in this case, which is the outcome of the hydrodynamic instability.  The magnetic energy is very 
time-dependent exhibiting irregular peaks and troughs. The drop of the magnetic energy corresponds to the breakdown of the LSCV to small scales but the magnetic field strength is re-established due to the re-formation of the LSCV. \cite{Lin2015} The cycle continues repeatedly but the periods are irregular because of the non-linear evolution.

\autoref{fig6:visualization_E3e-5Pm05} shows a snapshot of the vorticity $|\nabla \times \bm u|$ and the magnetic field $|\bm B|$ in the equatorial plane and in a meridional plane at $E=3\times 10^{-5}$, $P_o=-1.35\times 10^{-2}$,  $P_m=0.5$, $t/2\pi=1505$. This is a very representative snapshot when the LSCV are present.  In the equatorial plane, we can clearly see that the magnetic field is primarily generated in the region surrounding three dominant vortices. Note that the maximum magnetic magnitude appears between the vortices rather than at the centers of the vortices. The interactions (shearing and/or stretching) between vortices lead to strong strain surrounding the LSCV,  thus inducing magnetic fields. Since the LSCV are organized along the rotation axis of the fluid, the associated magnetic field also exhibits a columnar structure along the rotation axis of the fluid as we can see from the meridional plane in \autoref{fig6:visualization_E3e-5Pm05} (d). The distinctive role of the LSCV is   
 illustrated also by magnetic field lines in the fluid domain in \autoref{fig6:MagLines}. The field lines are stretched and twisted around three cyclones and show a columnar structure. 
 
Although the large scale magnetic fields are generated due to the LSCV in the bulk, the fields below the boundary ($r=1-\sqrt{E}$) are characterized by small scales in \autoref{fig6:BrBL}(a). The small scale fields in the boundary layer are much weaker than the large scale field associated with the LSCV in the bulk. We believe that the small scale fields are related with viscous boundary layer instabilities. \cite{Lorenzani2001,Kong2014} The magnetic fields are extended upward to outside of the fluid domain. Since the magnetic potential decays as $(1/r)^{l+1}$ outside the fluid domain, the field outside is mostly dipolar or quadrupolar. For example, \autoref{fig6:BrBL}(b) shows contours of $B_r$ on the surface $r=2$ (roughly corresponding to the Earth's surface if we assume that the boundary $r=1$ represents the core-mantle boundary). We observe a weak dipole field whose moment lies in the equatorial plane in this snapshot. However, the field structure varies in time (see Movie 2 in supplemental material \cite{Note1}). The exterior field can be either dipolar or quadrupolar with the orientation of the magnetic moment changing in time.

\autoref{fig6:spectra_E3e-5Pm05} shows energy spectra as a function of spherical harmonic degree $l$ at $t/2\pi=1505$, $E=3\times 10^{-5}$, $P_o=-1.35\times 10^{-2}$, $P_m=0.5$ (the same as \autoref{fig6:visualization_E3e-5Pm05}). The kinetic energy is dominated by the toroidal component with $l=1$, which corresponds to the solid body rotation response. The poloidal kinetic energy is dominated by an $l=2$ component. The maximum of the magnetic energy is at $l=4$ but the spectrum is almost flat for $l<10$. We note that the toroidal component of the magnetic energy is dominant compared to the poloidal component.       

\begin{figure}
\includegraphics[width=0.9\textwidth,clip,trim=0 3cm 0 0cm]{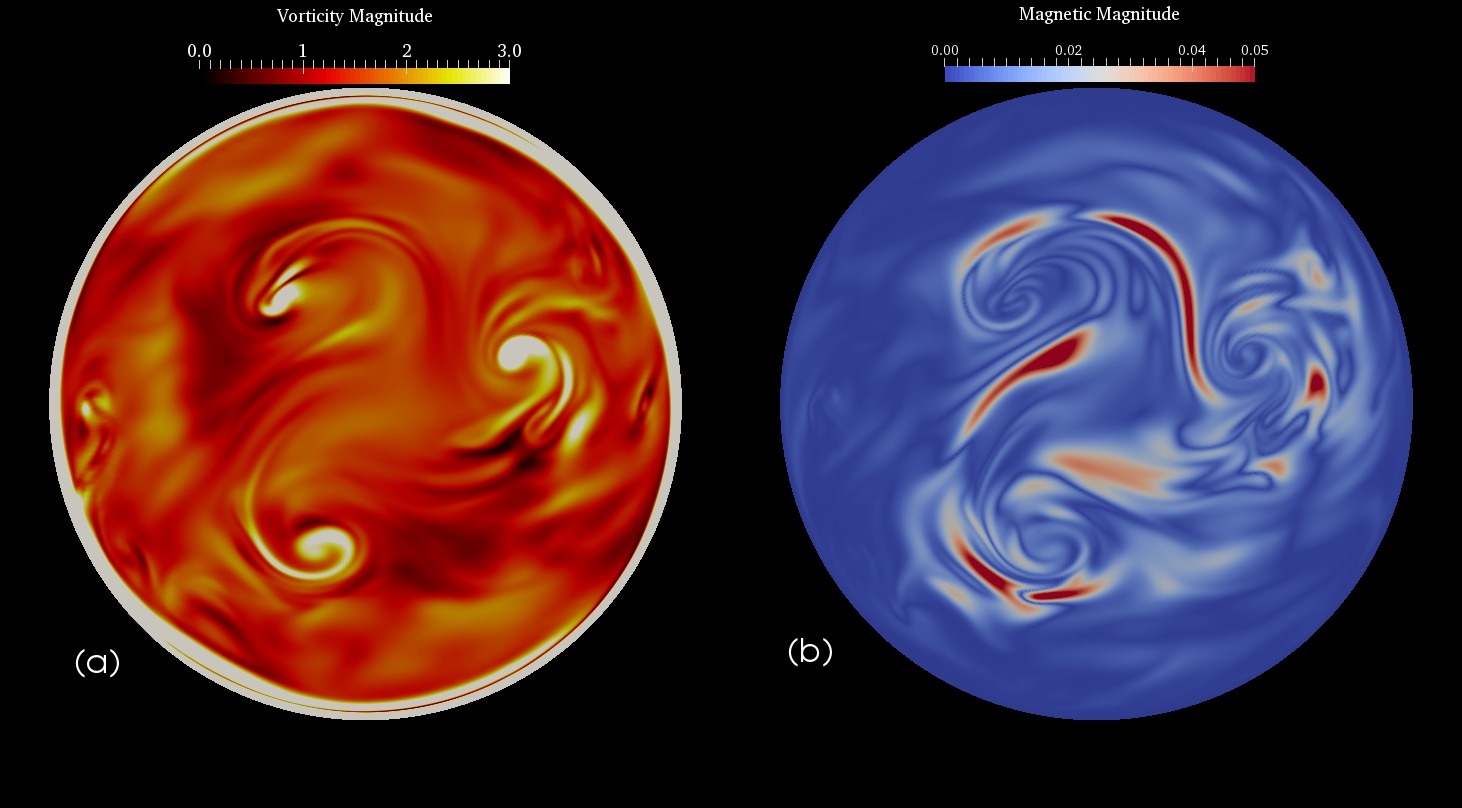}\\
\caption{Snapshot of the vorticity and the magnetic field at $E=6\times 10^{-5}$, $P_o=-2\times 10^{-2}$, $P_m=2$. (a) Vorticity $|\nabla \times \bm u|$ and (b) magnetic field strength $|\bm B|$ in the equatorial plane. 
}
\label{fig6:E6e-5Pm2}
\end{figure}

In \autoref{fig6:E6e-5Pm2}, we show another dynamo driven by LSCV at $E=6\times 10^{-5}$, $P_o=-2\times 10^{-2}$ and $P_m=2$. Qualitatively, this case is very similar with the one at $E=3\times 10^{-5}$, $P_o=-1.35\times 10^{-2}$, $P_m=0.5$. Again, we observe three dominant LSCV and the magnetic fields are mostly generated around the vortices. Indeed, the number of dominating LSCV, if they are formed, is three in all cases at different Ekman numbers and Poincar\'e numbers. This suggests that the length scale of the LSCV is independent of the Ekman number as reported in rotating convection simulations. \cite{Guervilly2015} Therefore, it is possible that similar large scale vortices can be formed in the liquid cores of planets, in which the Ekman number is extremely small.  

\begin{figure}
\includegraphics[width=0.95\textwidth]{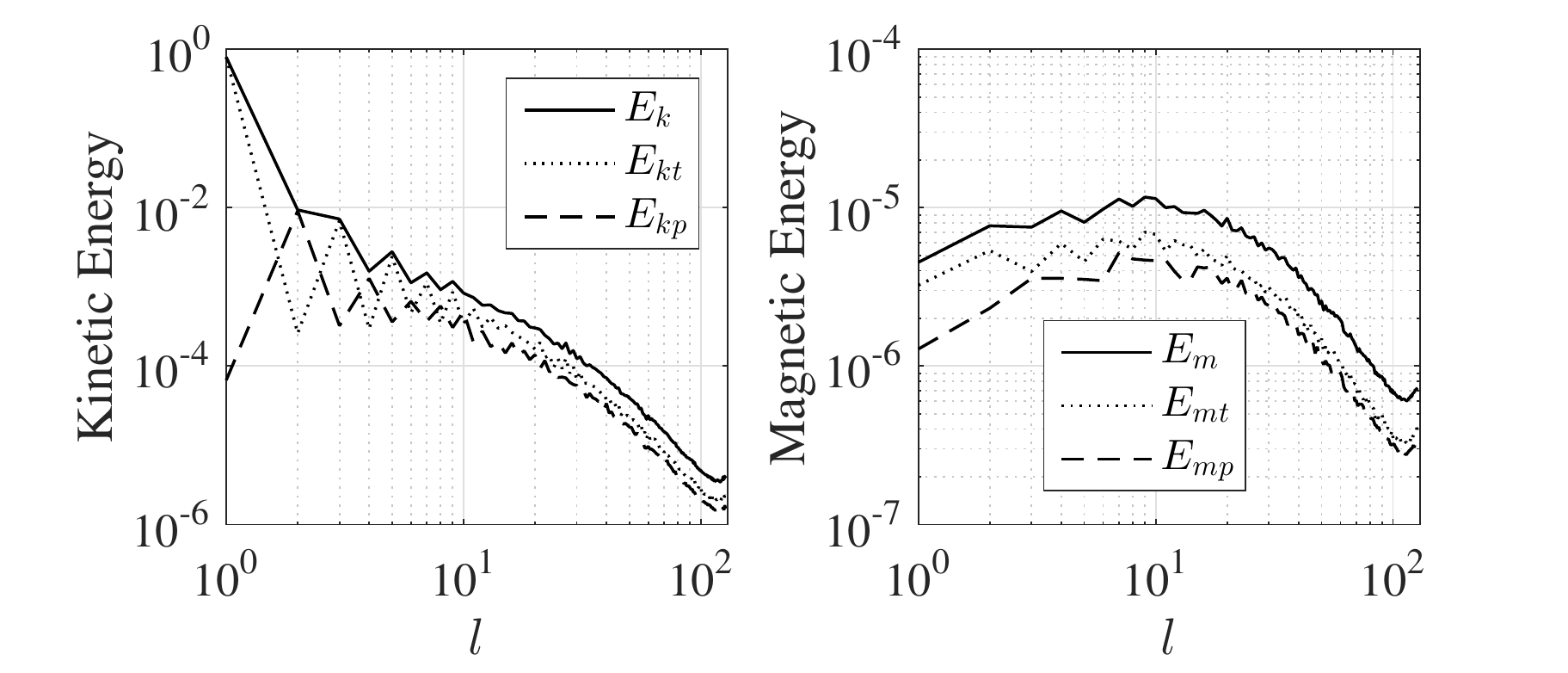} \\
(a)\hspace*{6cm}(b)
\caption{Snapshot of the energy spectra for the so-called turbulent ring dynamo at $E=1.0\times 10^{-4}$, $P_o=0.1$, $P_m=1$ and $\alpha_p=90^{\circ}$. (a) Kinetic energy spectra versus spherical harmonic degree $l$. (b) Magnetic energy spectra versus spherical harmonics degree $l$. Subscripts $t$ and $p$ represent the toroidal component and the poloidal component respectively.}
\label{fig6:spectra_E1e-4Pm1}
\end{figure}

However, there are conditions for the formation of the LSCV. First, we observe LSCV only at low Ekman numbers ($E<10^{-4}$), which means that the viscous effect should be much weaker than the rotational effect. Second, the precession rate should be large enough to trigger some instabilities but needs to be moderate to avoid strongly nonlinear effects. If the precession rate is sufficiently large ($|P_o|\gtrapprox0.1$), both numerical simulations \cite{Goto2014a, Lin2015} and laboratory experiments \cite{Goto2014} have shown small scale turbulence in the whole fluid volume and no formation of large scale vortices due to strong nonlinear effects compared to the rotational effect. Dynamo action in this strongly nonlinear regime has been reported from numerical simulations in a pressing sphere. \cite{Kida2011a} Our simulations have also reproduced this so-called turbulent ring dynamo using the same parameters as that of Ref \onlinecite{Kida2011a}, i.e. $E=1.0 \times 10^{-4}$, $P_o=0.1$, $P_m=1.0$ and $\alpha_p=90^{\circ}$ ($\bm{\Omega_o}$ and $\bm{\Omega_p}$ are orthogonal). The field structure of this dynamo has been shown in detail in Ref \onlinecite{Kida2011a}. Here we shown only the energy spectra for this case in \autoref{fig6:spectra_E1e-4Pm1} , which were not presented in Ref \onlinecite{Kida2011a}. The total kinetic energy is again dominated by the toroidal component with $l=1$. The main contribution to the poloidal kinetic energy is from $l=2$ component. The magnetic spectra clearly show smaller scale fields compared to dynamos driven by LSCV. We see the broadband magnetic energy spectra with a maximum around $l=10$. In addition, the toroidal component and poloidal component  make comparable contributions to the total magnetic energy in this case.           

\section{Discussion}\label{sec:Diss}
Based on previous hydrodynamic simulations, we have shown precession driven dynamos in different flow regimes. In the laminar regime, dynamo action operates mainly in a thin layer beneath the boundary since the bulk fluid is nearly a solid body rotation.    
Our simulations at lower Ekman numbers than previous studies clearly show that precession driven laminar dynamos are more difficult to obtain at low Ekman numbers, as has been pointed out previously. \cite{Tilgner2005} The main result of the present study is that we have demonstrated magnetic field generation by large scale vortices in a precessing sphere, which has not been explored previously. Small scale flows driven by precessional instabilities can merge into large scale cyclonic vortices  due to the effect of rapid rotation. \cite{Mouhali2012,Lin2015} These vortices are elongated along the rotation axis of the fluid and the length scale of the vortices is independent of the Ekman and Poincar\'e numbers in the parameter ranges that we have explored. Due to the shearing between the vortices, magnetic fields are mainly generated in the region surrounding the cyclonic vortices with the fields also showing a columnar structure (\autoref{fig6:MagLines}). It seems that the presence of the large scale vortices facilitates the onset of dynamo action, i.e. by lowering the critical magnetic Prandtl number. This mechanism makes it possible to sustain dynamos at low Ekman numbers, Poincar\'e numbers and magnetic Prandtl numbers, which we believe moves towards the parameters regime of planets. In particular, we have demonstrated dynamo action in the regime of magnetic Prandtl number $P_m<1$, in which the diffusivities have the correct hierarchy with respect to the fact that $P_m \approx  10^{-5}$ for liquid metals and for the liquid cores of planets. In addition, it seems that the critical magnetic Prandtl numbers for dynamos drops as one  decreases the Ekman number if the large scale cyclonic vortices are formed (\autoref{tab:models}), although more generic scaling laws are required. Nevertheless, these observations are promising with respect to further investigations and planetary applications.

In planetary settings, one would question whether precession driven flow is laminar or unstable in liquid cores. We have addressed this question for the Earth and the Moon in Ref \onlinecite{Lin2015}. Here we briefly recall some conclusions. 
Considering the luni-solar precession of the Earth with a period of around 26,000 years, the criterion for all known instability mechanisms \cite{Kerswell1993, Lorenzani2001, Lorenzani2003, Lin2015} of the precession-driven flow is not matched. This suggests that the precession-driven flow in the Earth's fluid core may be stable. Hence we do not expect that the precession plays a significant role in the generation of the Earth's magnetic field.    
However, there are still several uncertainties concerning this conclusion due to the possible effects of the solid inner core \cite{Tilgner1999a} and interactions between precession and convection. \cite{Wei2013}
For the case of the Moon, we have estimated that the growth rate of precessional instability due to the conical shear layer is two orders of magnitude larger than the viscous decay rate, \cite{Lin2015} suggesting very complex flows in the lunar core driven by the 18.6 yr precession of the moon. Therefore, a lunar dynamo driven by precession is possible during the evolution history of the Moon, \cite{Dwyer2011} particularly if the large scale vortices are formed due to the precessional instability. However, the Moon does not have an observable internal magnetic field generated by dynamo action at present time. Note that not all of the power deposited by the precession is available to sustain a lunar dynamo. \cite{Dwyer2011} There is also a threshold power requirement to maintain the lunar core in a well-mixed adiabatic state, which is not matched at present day. \cite{Williams2001,Dwyer2011}

Although we have made great efforts to push towards the parameter regime of planets, our simulations are still far away from the realistic parameter regime, and there is little prospect of approaching the realistic parameters  which require considerable computational resources. Therefore, it would be helpful in future to extract some generic scaling laws from numerical models as in the studies of convection driven dynamos. \cite{Christensen2006,Stelzer2013} On the other hand, laboratory experiments with liquid metals can reach more extreme parameters, which would significantly compensate for the limitations of numerical models. A liquid 
sodium experiment of a precessing cylinder with the height and diameter of 2 meters is under construction in Dresden, Germany,  \cite{Stefani2015} which is expected to provide new insights on precession driven dynamos.

\begin{acknowledgments}
We gratefully thank R. Hollerbach, S. Vantieghem and D. C\'ebron for useful discussions, and three anonymous referees for constructive comments. This study was supported by the ERC Grant No. 247303 (MFECE) at ETH Zurich. YL acknowledges the Swiss National Science Foundation for a PostDoc Mobility fellowship at the University of Cambridge. PM acknowledges the support of NSF CSEDI program through award EAR-1067944. Simulations were run on Swiss National Supercomputing Center (CSCS) under the project s369. The authors would like to thank Jean Favre at CSCS for assistance with visualizations.         
\end{acknowledgments}


%

\end{document}